\documentclass[preprint,prd,nofootinbib,tightenlines,amsmath,amssymb]{revtex4}
\usepackage[dvipdfmx]{graphicx}
\usepackage{bm}
\usepackage{amsmath}
\usepackage{hhline}
\usepackage{color}
\usepackage{xcolor}
\usepackage{slashed}
\usepackage{relsize}

\newcommand{\Eq}{&=&}

\begin{document}

\vspace*{3em}

\preprint{TU-1093, IPMU19-0122}%
\preprint{}%

\title{Hidden Monopole Dark Matter via Axion Portal and its Implications for 
Direct Detection Searches, Beam-Dump Experiments, and the $H_0$ Tension
} 

\author{
Ryuji Daido,$^{1}$\footnote[3]{daido@tuhep.phys.tohoku.ac.jp}
Shu-Yu Ho,$^{1}$\footnote[1]{ho.shu-yu.q5@dc.tohoku.ac.jp}
Fuminobu Takahashi,$^{1,2}$\footnote[2]{fumi@tohoku.ac.jp}
}
\affiliation{
${}^{1}$Department of Physics, Tohoku University, Sendai, Miyagi 980-8578, Japan \vspace{3pt} \\
${}^{2}$Kavli Institute for the Physics and Mathematics of the Universe (Kavli IPMU), UTIAS, WPI, The University of Tokyo, Kashiwa, Chiba 277-8568, Japan\vspace{3ex}}


\begin{abstract}
Hidden monopole is a plausible dark matter candidate due to its stability, but its direct experimental search is extremely difficult due to feeble interactions with the standard model particles in the minimal form. Then, we introduce an axion, $a$, connecting the hidden monopole and the standard model particles and examine the current limits and future prospects of direct dark matter searches and beam-dump experiments. We find two parameter regions around $m_a = {\cal O}(10)$\,MeV, $f_a = {\cal O}(10^{5})$\,GeV and $m_a = {\cal O}(100)$\,MeV, $f_a = {\cal O}(10^{4})$\,GeV where monopole dark matter and the axion are respectively within the reach of the future experiments such as PICO-500 and SHiP. We also note that the hidden photons mainly produced by the axion decay contribute to dark radiation with $\Delta N_{\rm eff} \simeq 0.6$ which can relax the $H_0$ tension.
\end{abstract}

\maketitle

\section{Introduction}\label{sec:1}
Dark matter (DM) is ubiquitous in the universe, making up about a quarter of the present energy density. It is also known to play 
a crucial role in structure formation. However, what DM is composed of remains one of the great mysteries in particle physics and cosmology. 

One of the peculiar properties of DM is its stability. Its lifetime must be at least an order of magnitude longer than the current age of the universe~\cite{Enqvist:2015ara,Enqvist:2019tsa}. The stability of DM may be ensured by  symmetry. For instance, in the WIMP scenario, its stability is usually achieved by imposing $Z_2 $ symmetries such as $R$-parity and KK-parity. It is also possible that the stability of DM is due to an unbroken gauge symmetry (see Ref.~\cite{Agrawal:2016quu} and references therein). One such example is a hidden magnetic monopole, which is the main focus of this paper.

A monopole is a topological defect which appears associated with the spontaneous breaking of a gauge symmetry ${\cal G}$ down to ${\cal H}$ with a non-trivial $\pi_2({\cal G}/{\cal H})$. The simplest example is the 't Hooft-Polyakov monopole~\cite{tHooft:1974kcl,Polyakov:1974ek} with ${\cal G}=$ SU(2) and ${\cal H}=$ U(1), and it is the magnetic charge under the U(1) gauge symmetry that makes the monopole absolutely stable. If there is a hidden sector which includes such a gauge symmetry and its subsequent spontaneous breaking allows the monopole solution, then hidden monopoles are necessarily produced at the  phase transition, and they can contribute to DM.\footnote{In the following, a monopole refers to a hidden monopole, not the monopole with the ordinary magnetic charge nor a GUT monopole.} The hidden monopole DM was studied in detail in Refs.\,\cite{Murayama:2009nj,Baek:2013dwa,Khoze:2014woa}. Through the Witten effect~\cite{Witten:1979ey}, the monopole DM can suppress the abundance and the isocurvature perturbations of the QCD axion~\cite{Kawasaki:2015lpf,Nomura:2015xil,Kawasaki:2017xwt,Sato:2018nqy}.

The monopole is a promising candidate for DM, but it is formidable to search directly by experiments because its interactions to the standard model (SM) particles are feeble. In fact, in the minimal setup, it is completely decoupled from the SM sector at a renormalizable level. The setup can be extended to couple the monopole to the SM sector via the Higgs, vector, or axion portal coupling. In fact, the Higgs portal coupling was assumed in Refs.\,\cite{Baek:2013dwa,Khoze:2014woa} in order to keep the hidden sector in thermal equilibrium with the SM plasma. The monopole can interact with the nucleon via the Higgs portal, but the expected cross section is much smaller than any direct DM search experiments in the foreseeable future. On the other hand, the vector portal coupling induces a small electric charge for the hidden monopole. The mini-charged DM is strictly constrained by various experiments and observations~\cite{Tanabashi:2018oca}. We will come back to this possibility later in this paper. Our main interest is the axion portal coupling. As described below, the monopole DM scatters off nucleons with a cross section within the range of future experiments such as PICO-500 for a certain parameter region. Furthermore, axions can be probed by beam-dump experiments such as SHiP. As we shall see later, we find two parameter regions with the axion mass and decay constant given by $m_a = {\cal O}(10)$\,MeV, $f_a = {\cal O}(10^{5})$\,GeV and $m_a = {\cal O}(100)$\,MeV, $f_a = {\cal O}(10^{4})$\,GeV, where both monopole DM and the axion are within the reach of the direct DM searches and SHiP.

The hidden sector contains a massless hidden photon which contributes to dark radiation. The amount of dark radiation is customarily expressed in terms of the extra effective neutrino species, $\Delta N_{\rm eff}$. Due to the axion portal coupling, the hidden photon abundance is mainly determined by the axion decay when the cosmic temperature becomes comparable to the axion mass. In particular, for the above parameter space where both monopole DM and the axion are within the experimental reach, the amount of dark radiation is predicted to be $\Delta N_{\rm eff} \simeq 0.6$. We note that such an amount of dark radiation can significantly relax the $H_0$ tension~\cite{Riess:2019cxk}.

The rest of this paper is organized as follows. In Sec.\,\ref{sec:2}, we briefly review the monopole DM and the Witten effect. In Sec.\,\ref{sec:3}, we introduce the axion portal coupling and solve for the axion profile around a monopole. Using the axion profile, we estimate the scattering cross-section of the hidden monopole with nucleons and discuss its implications for future experiments. In Sec.\,\ref{sec:4}, we discuss the current limits and future prospects of various  experiments including direct DM search experiments and the beam-dump experiments. The last section is devoted to discussion and conclusions.

\section{Monopole dark matter}\label{sec:2}
\subsection{'t Hooft-Polyakov Monopoles}\label{subsec:1}
't Hooft and Polyakov showed that a magnetic monopole can arise when a non-Abelian gauge symmetry is spontaneously broken via the Higgs mechanism~\cite{tHooft:1974kcl,Polyakov:1974ek}. The simplest possibility is an SU(2)$_{\rm H}$ gauge theory with gauge fields $\boldsymbol{A}_{\rm H}^\mu = (A_{\rm H 1}^{\mu},A_{\rm H 2}^{\mu},A_{\rm H 3}^{\mu})$ and an isovector scalar field $\boldsymbol{\phi} = (\phi_1,\phi_2,\phi_3)$. The Lagrangian density is given by
\begin{eqnarray}
{\cal L}_{\rm H} 
\,=\, 
-\frac{1}{4}  \boldsymbol{F}_{\rm H}^{\mu\nu} \cdot \boldsymbol{F}^{}_{{\rm H}\mu\nu}
+ \frac{1}{2} {\cal D}^\mu  \boldsymbol{\phi} \cdot {\cal D}_\mu \boldsymbol{\phi}\, 
- {\cal V}(\phi) 
~,
\end{eqnarray}
where $\boldsymbol{F}_{\rm H}^{\mu\nu} = \partial^\mu \boldsymbol{A}_{\rm H}^\nu - \partial^\nu \boldsymbol{A}_{\rm H}^\mu + e^{}_{\rm H} \boldsymbol{A}_{\rm H}^\mu \times \boldsymbol{A}_{\rm H}^\nu$ is the field strength tensor of the SU(2)$_{\rm H}$ gauge fields with $e^{}_{\rm H}>0$ being the hidden gauge coupling constant, and ${\cal D}^\mu \boldsymbol{\phi} = \partial^\mu \boldsymbol{\phi} + e^{}_{\rm H} \boldsymbol{A}_{\rm H}^\mu \times \boldsymbol{\phi}$ is the covariant derivative.\footnote{Note that $\cdot$ and $\times$ are the dot and cross products in the isospin space.} The potential of the isovector scalar field is
\begin{eqnarray}
{\cal V}(\phi) 
\,=\, 
\frac{1}{4} \lambda_\phi \big( \phi^2 - \upsilon^2_{\rm H} \big)^2 
~,\quad 
\phi\,=\, \sqrt{\boldsymbol{\phi} \cdot \boldsymbol{\phi}} 
~,
\end{eqnarray}
where $\lambda_\phi > 0$ is the quartic coupling. The potential is minimized at $ \langle \phi \rangle = \upsilon^{}_{\rm H}$, where SU(2)$_{\rm H}$ is broken down to an unbroken U(1)$_{\rm H}$.

Now if we take $\langle \boldsymbol{\phi} \rangle = (0,0,\upsilon^{}_{\rm H})$, the unbroken U(1)$_{\rm H}$ symmetry corresponds to rotations about the $3$-axis, and the hidden photon is identified with $A_{\rm H 3}^{\mu}$. Expanding the Lagrangian density around the vacuum state, we then obtain a massless hidden photon $\gamma^{}_{\rm H}$, two massive hidden gauge bosons $W_{\rm H}^\pm$, and a massive neutral Higgs field $\varphi$.  In addition, this theory allows a static field configuration with finite energy due to the nontrivial $\pi_2({\rm SU(2)_H}/{\rm U(1)_H}) = Z$. Such a topological soliton with nonzero winding number possesses a magnetic charge, and is therefore identified with the magnetic monopole $\text{M}$. The hidden magnetic charge of the monopole with a unit winding number is $Q_{\rm M} = 4 \pi/e^{}_{\rm H}$.

We summarize their masses and electric and magnetic charges in Table~\ref{tab:particle_spectrum}. In particular, the mass of the hidden monopole is bounded from below by the Bogomol'nyi bound\,\cite{Bogomolny:1975de}, $m^{}_{\rm M} \geq m^{}_{W_{\rm H}}/\alpha^{}_{\rm H}\,{\cal F}(\lambda_\phi/\alpha^{}_{\rm H})$, where ${\cal F}$ is a monotonically increasing function with ${\cal F}(0)=1$ and ${\cal F}(\infty) \simeq 1.787$\,\cite{Preskill:1984gd}. Throughout this paper, we will adopt the hidden monopole mass saturating the Bogomol'nyi limit ($\lambda_\phi\to 0$) for simplicity following Ref.\,\cite{Khoze:2014woa}.
The electric charge of the hidden monopole due to the Witten effect will be explained in the next subsection.

\renewcommand{\arraystretch}{1.2}
\begin{table}[h!]\
\vspace{-0.5cm}
\caption{The masses, electric and magnetic charges of the particles in the hidden sector, where $\alpha^{}_{\rm H} = e^2_{\rm H}/(4\pi)$ is the hidden fine structure constant. Here we take the Bogomol'nyi limit for the hidden monopole mass and its hidden electric charge is due to the Witten effect.}
\vspace{0.3cm}
\centering 
\begin{tabular}{|c|c|c|c|}
\hline
~\text{Particle}~   & ~\text{Mass}~  & ~\text{Hidden electric charge}~ & ~\text{Hidden magnetic charge}~ \\
\hline
$\gamma^{}_{\rm H}$  & ~0~  & ~0~  & ~0~ \\[0.1cm]
$\varphi$   & ~$m_\varphi = \sqrt{2\lambda_\phi} \, \upsilon^{}_{\rm H}$~ & ~0~  & ~0~ \\[0.1cm]
$W_{\rm H}^\pm$  & ~$m^{}_{W_{\rm H}} = \sqrt{4\pi \alpha^{}_{\rm H}} \,\upsilon^{}_{\rm H}$~     & ~$Q_{\rm E} = \pm e^{}_{\rm H}$~ & ~0~ \\[0.1cm]
$\text{M} (\overline{\text{M}})$  & ~$m^{}_{\rm M} = \sqrt{4\pi / \alpha^{}_{\rm H}} \, \upsilon^{}_{\rm H}$~  & ~$Q_{\rm E} = \pm e^{}_{\rm H}\theta^{}_{\rm H}/(2\pi)$~ & ~$Q_{\rm M} = \pm 4\pi/e^{}_{\rm H}$~\\[0.1cm]
\hline
\end{tabular}
\vspace{0.2cm}
\label{tab:particle_spectrum}
\end{table}

\subsection{The Witten Effect}\label{subsec:2}
To see how the hidden monopole acquires a nonzero electric charge, let us briefly review the consequence of adding the following P- and CP-violating $\theta_{\rm H}$-term in the hidden sector,
\begin{eqnarray}
\label{witteneffect}
{\cal L}_\theta 
\,=\, 
\theta_{\rm H} \frac{e_{\rm H}^2}{32\pi^2} F_{\rm H}^{\mu\nu} \widetilde{F}^{}_{{\rm H}\mu\nu} 
~,
\end{eqnarray}
where $\theta_{\rm H}$ is a constant,  $F_{\rm H}^{\mu\nu} = \partial^\mu A_{\rm H3}^\nu - \partial^\nu A_{\rm H3}^\mu$ is the field strength tensor of the hidden photon, and $\widetilde{F}_{\rm H}^{\mu\nu}$ is its dual tensor. We will promote $\theta_{\rm H}$ to a dynamical variable, the axion, in the next section. Since the $\theta_{\rm H}$-term can be rewritten as a total derivative, it does not affect the equations of motion, and one usually dismisses such a term. However, in the presence of the monopole, it has a physical effect. In other words, the monopole acquires an electric charge proportional to $\theta_{\rm H}$, in addition to the magnetic charge. This is known as the Witten effect\ \cite{Witten:1979ey}. As a result, the monoploe becomes a dyon\,\cite{Schwinger:1969ib}.
 
To derive the explicit form of the electric charge of the hidden monopole, let us start with the equations of motion of the hidden electromagnetism, 
\begin{eqnarray}
\partial_\mu 
F_{\rm H}^{\mu\nu}
\,=\,
\frac{e_{\rm H}^2}{8\pi^2} 
\partial_\mu \big(\theta_{\rm H} \widetilde{F}_{\rm H}^{\mu\nu}\big)
~,\quad
\partial_\mu \widetilde{F}^{\mu \nu} &= j_{\rm M}^\nu
~,
\end{eqnarray}
where $ j_{\rm M}^\mu = (\rho_{\rm M}, \boldsymbol{J}_{\rm M})$ is the magnetic 4-current density. Noting that the electric and magnetic fields appear in the field strength as $F_{\rm H}^{j0} = E_{\rm H}^j$ and $F_{\rm H}^{jk} = -{}^{}\epsilon_{jkl}B_{\rm H}^l$, we obtain the following modified Maxwell's equations,
\begin{eqnarray}
\label{hiddenEM}
\nabla \cdot \boldsymbol{E}_{\rm H} 
\,=\,
\frac{e_{\rm H}^2}{8\pi^2} 
\nabla \cdot \big(\theta_{\rm H} \boldsymbol{B}_{\rm H}\big)
~,\quad
\nabla \cdot  \boldsymbol{B}_{\rm H} & = \rho_{\rm M}
~.
\end{eqnarray}

Suppose that there exists a monopole with the magnetic charge $Q_{\rm M}$ sitting at the origin. When measured far from the origin, the magnetic field induced by the monopole is
\begin{eqnarray}
\boldsymbol{B}_{\rm H}(r) 
\,=\, 
\frac{Q_{\rm M}}{4\pi} \frac{\hat{\boldsymbol{r}}}{r^2} 
~,
\end{eqnarray}
where $r$ is the distance away from the origin and $\hat{\boldsymbol{r}}$ denotes the unit vector along the radial direction. Here we assume that the core radius of the monopole is negligibly small compared to $r$. Using (\ref{hiddenEM}), one can see that a nonzero hidden electric field is also generated around the monopole at a sufficiently large $r$,
\begin{eqnarray}\label{Efield}
\boldsymbol{E}_{\rm H}(r)
\,=\, 
\frac{Q_{\rm E}}{4\pi} \frac{\hat{\boldsymbol{r}}}{r^2} 
\end{eqnarray}
with
\begin{eqnarray}
Q_{\rm E} 
\,=\, 
\frac{e_{\rm H}^2}{8\pi^2} \theta_{\rm H} Q_{\rm M}
\,=\, 
\frac{e_{\rm H}\theta_{\rm H}}{2\pi}
~.
\end{eqnarray}
Thus, the hidden monopole acquires a nonzero hidden electric charge (see Table~\ref{tab:particle_spectrum}). Note that anti-monopoles have the opposite hidden electric charges. 

One can extend this result by promoting $\theta_{\rm H}$ to an axion field and show that the electric charge of the monopole is similarly induced by the Witten effect~\cite{Fischler:1983sc}. The difference is that the electric charge spreads out in space, and the electric charge contained inside a sphere with a radius $r$ is given by $Q_{\rm E}(r) = e_{\rm H}\theta(r)/(2\pi)$, following
the axion field configuration around the monopole. We will study the axion configuration around the monopole in Sec.~\ref{sec:3}.

\subsection{Relic Abundance}\label{subsec:3}

In the 't Hooft-Polyakov monopole model, both massive gauge bosons and monopoles are stable and contribute to DM. This is because they carry  electric and/or magnetic charges under unbroken U(1)$_{\rm H}$ symmetry and there are no other lighter charged particles. 

The hidden sector is assumed to be in thermal equilibrium with the SM sector through some portal interactions. Then, the massive gauge bosons are in thermal equilibrium and their abundance is fixed when the annihilation processes such as $W^+_{\rm H}W^-_{\rm H} \leftrightarrow \gamma^{}_{\rm H}\gamma^{}_{\rm H}, \varphi\varphi$ are decoupled as in the usual freeze-out scenario. On the other hand, monopoles are produced at the phase transition, and their abundance depends on the order of the phase transition.

Throughout this paper, we assume that the universe experiences a second-order phase transition at the critical temperature $T_{\rm c}$ in the hidden sector so that the monopole abundance is determined by the Kibble-Zurek mechanism~\cite{Kibble:1976sj,Zurek:1985qw}.\footnote{In the case of the first-order phase transition, the monopole mass is
predicted to be at intermediate scales since their number density is much smaller~\cite{Khoze:2014woa}.} At high temperatures, $T \gg T_{\rm c}$, the SU(2)$_{\rm H}$ symmetry is restored, and the vacuum expectation value of the isovector scalar field vanishes everywhere in the universe. When $T \lesssim T_{\rm c}$, the SU(2)$_{\rm H}$ symmetry is spontaneously broken, then the value of $\langle \phi \rangle$ changes from 0 to $\upsilon^{}_{\rm H}$, and a domain structure appears. The size of each domain is characterized by the correlation length, $\zeta$, and the direction of $\boldsymbol{\phi}$ in each domain is randomly chosen. Toward the phase transition, both the correlation length and the relaxation time scale increase proportional to $|T-T_{\rm c}|^{-\nu}$ with $\nu$ being the critical exponent. The classical value of the critical exponent is $\nu = 1/2$. This increase can be understood by noting that the dynamical time scale is determined by the effective mass of the Higgs, which becomes smaller and approaches zero toward the critical point. At a certain point, the relaxation time becomes equal to the time left for the universe to reach the critical point, beyond which the typical size of the domains is frozen. Then, approximately one monopole is produced in the volume of order $\zeta^3$. The monopole abundance from the Kibble-Zurek mechanism is thus estimated as~\cite{Zurek:1985qw,Murayama:2009nj}
\begin{eqnarray}\label{relicM}
\Omega_{\rm M} h^2 \,\simeq\, 
1.5 \times 10^{9} 
\bigg(\frac{m_{\rm M}}{1\,\text{TeV}}\bigg)
\bigg(30\frac{T_{\rm c}}{M_{\rm Pl}}\bigg)^{3\nu/(1+\nu)} ~,
\end{eqnarray}
where $h \simeq 0.7$ is the reduced Hubble parameter, and $M_{\rm Pl} \simeq 1.22 \times 10^{19}$\,GeV is the Planck mass. Notice that the monopole-anti-monopole annihilation after the phase transition is taken into account in deriving \eqref{relicM}. If one takes $m_{\rm M}/T_{\rm c} \simeq m_{\rm M}/\upsilon^{}_{\rm H} = \sqrt{4\pi/\alpha^{}_{\rm H}} \simeq {\cal O}(10)$, then the hidden monopole mass should be about several hundred TeV to sub-PeV scale in order to explain the observed DM density~\cite{Murayama:2009nj}. 

The monopole DM has sizable self-interactions because it has both magnetic and electric charges. Although there are various observational constraints on self-interacting DM, such heavy monopole DM is considered to escape the restriction~\cite{Khoze:2014woa,Agrawal:2016quu}. In our numerical calculation, we will pick up $\alpha^{}_{\rm H} \simeq 0.73$, $m_{\rm M}\simeq 216 \,\text{TeV}$, and $\nu = 0.5$ as a benchmark point, for which around 35\% of the DM is contributed by the hidden monopole~\cite{Khoze:2014woa}. Note that, although the precise fraction of the monopole DM is subject to various uncertainties such as the critical exponent and the presence of hidden fermions that are necessary for the interaction of the axion to be introduced in the next section, they do not change our results by more than an order of magnitude at least in the parameter region of our interest.

\section{Spin-dependent monopole-nucleon interactions through the axion portal}\label{sec:3}
In this section, we will demonstrate how to calculate the spin-dependent cross-section of the hidden monopole DM scattering off a nucleon by introducing the axion portal. Such calculation may look intractable as the monopole is a composite object and it is not clear how to express its interaction with the axion and the SM particles in the Lagrangian. In the following, we shall first evaluate the axion configuration around the monopole due to the Witten effect. If the axion has Yukawa-like interactions with the SM fermions, the monopole surrounded by the axion can scatter off a nucleon. Then, its leading-order cross section can be estimated by treating the axion configuration as an external field. This enables us to study the current limits and future prospects of various direct DM search experiments.

\subsection{Axion portal}
To find out the axion profile around the hidden monopole, let us consider the following Lagrangian density,
\begin{eqnarray}\label{axion_portal}
{\cal L}_a
\,=\, 
-\frac{1}{4} F_{\rm H}^{\mu\nu} F^{}_{{\rm H}\mu\nu} 
+\frac{1}{2} \partial^\mu a \partial_\mu a
-\frac{1}{2} m_a^2 f_{\rm H}^2 \bigg(\frac{a}{f^{}_{\rm H}}-\theta_0\bigg)^2
+ \frac{e_{\rm H}^2}{32\pi^2} \frac{a}{f_{\rm H}}  F_{\rm H}^{\mu\nu} \widetilde{F}^{}_{{\rm H}\mu\nu}
~,
\end{eqnarray}
where $a = a(r)$ is the axion field with $f^{}_{\rm H}$ being the decay constant of the axion. If the shift symmetry of the axion is linearly realized as global U(1)$_{\rm PQ}$ symmetry in the UV completion, it resides in the phase of a complex scalar, $S$.
Then, the decay constant is considered to be of order the vacuum expectation value of $S$ in a simple setup. Also, the axion coupling to the hidden photons is generated if $S$ has a coupling to hidden fermions charged under SU(2)$_{\rm H}$. The mass of the hidden fermions is also expected to be of the same order.\footnote{The massive gauge boson $W_{\rm H}^\pm$ may decay into those fermions if kinematically allowed. For the parameter region of our interest, the mass of hidden fermions is not very different from $W_{\rm H}^\pm$, and so, this does not change our result significantly.} In Eq.~(\ref{axion_portal}), we have introduced the axion mass term. This may originate from another hidden gauge theory SU$({\rm N})_{\rm H'}$ with gauge fields $G_{\rm H'}$, in which $\theta_0$ is the coefficient of the theta term, $\theta_0 G_{\rm H'}\widetilde{G}_{{\rm H}'}$, and the mass term is obtained by expanding the potential around $\theta_0$. The expansion may break down when $\theta_0 \gtrsim 1$ as $a/f^{}_{\rm H}$ changes its value more than ${\cal O}(1)$ around the monopole. We, however, neglect the possible deviation from the quadratic potential because the calculation of the scattering cross section is essentially determined by the tail of the axion configuration where the axion potential can be well approximated by the quadratic form. Alternatively, one can simply break the shift symmetry of the axion just by introducing such a mass term. 

We suppose that, in the absence of monopoles, or at a sufficiently large distance from monopoles, the axion is stabilized at $\theta_0$. Now, let us place a monopole at the origin and derive the static field configuration of the axion around it. 
Due to the Witten effect, the monopole is surrounded by an electric charge, and the total amount of the electric charge is given by $e_{\rm H} \theta_0/2\pi$, which is independent of the axion decay constant. The electric field becomes stronger as the axion approaches the monopole. As a result, the axion field value becomes close to zero to suppress the energy of the electric field. Thus, the axion configuration is determined by the balance between the gradient energy of the axion and the energy stored in the electric field around the monopole.

To derive the precise axion distribution around the monopole, one needs to minimize the Hamiltonian of the axion-monopole system given by
\begin{eqnarray}
H_{a-{\rm M}} 
\Eq
\mathop{\mathlarger{\int}} \hspace{-0.1cm} d^3x
\bigg[
\frac{1}{2} \dot{\theta}^2+
\frac{1}{2} f_{\rm H}^2 (\nabla \theta)^2 
+\frac{1}{2} m_a^2 f_{\rm H}^2 (\theta-\theta_0)^2
+\frac{1}{2}|\boldsymbol{E}_{\rm H}|^2
+\frac{1}{2}|\boldsymbol{B}_{\rm H}|^2
\bigg]
\nonumber\\[0.2cm]
\Eq
2\pi f_{\rm H}^2 \mathop{\mathlarger{\,\int}_{r_{\rm c}}^\infty} \hspace{-0.1cm} dr
\bigg[
\bigg(r\frac{d\theta(r)}{dr}\bigg)^2 
+m_a^2 r^2 \big(\theta(r)-\theta_0\big)^2
+\frac{r_0^2}{r^2} \theta(r)^2
\bigg]+{\rm \,const.}
~,
\end{eqnarray}
where we have defined a dimensionless axion field, $\theta \equiv a/f_{\rm H}$, and we have used Eqs.\,\eqref{Efield} and \eqref{axion_portal} and omitted the kinetic energy in the second line as we are interested in the static field configuration. We also drop the energy density of the hidden magnetic field since it does not contribute to the equation of motion of the axion.\footnote{Note that the theta term $\theta F\widetilde{F}$ in Eq.\,\eqref{witteneffect} does not contribute the Hamiltonian density even if now $\theta$ is spacetime dependent. The easiest way to  prove this  is to consider the following action 
\begin{eqnarray}
S_\theta 
\,=\, \int d^4 x \, \sqrt{-g} \, \theta(x) F_{\mu\nu} \widetilde{F}^{\mu\nu}
\,=\, \tfrac{1}{2} \epsilon^{\mu\nu\alpha\beta} \int d^4 x \, \theta(x)
F_{\mu\nu} F_{\alpha\beta}~,
\nonumber
\end{eqnarray}
where $\widetilde{F}^{\mu\nu} = \tfrac{1}{2}\epsilon^{\mu\nu\alpha\beta}F_{\alpha\beta}/\sqrt{-g}$. Therefore, $T^{\mu\nu}_\theta = \delta S_\theta/\delta g_{\mu\nu} = 0$.} We have assumed spherical symmetry of the solution in the second line, and 
$r_0$ is defined as
\begin{eqnarray}
r_0 \,\equiv\, \frac{e^{}_{\rm H}}{8\pi^2 f^{}_{\rm H}}
~.
\end{eqnarray}
The lower end of the integration is set by the core radius of the hidden monopole, $r_{\rm c} \simeq m^{-1}_{W_{\rm H}}$, inside which the original SU(2)$_{\rm H}$ gauge symmetry is restored. In the following analysis, we will assume that the core radius is negligibly small. This is justified because the typical momentum transfer in the monopole-nucleon scattering is much smaller than the mass of the vector boson. Also, the axion profile outside the core is considered to be insensitive to $r_{\rm c}$, since the electric field outside is simply determined by the Gauss's law, and  it does not depend on $r_{\rm c}$. 

\begin{figure}[t!]
\centering
\includegraphics[scale=0.65]{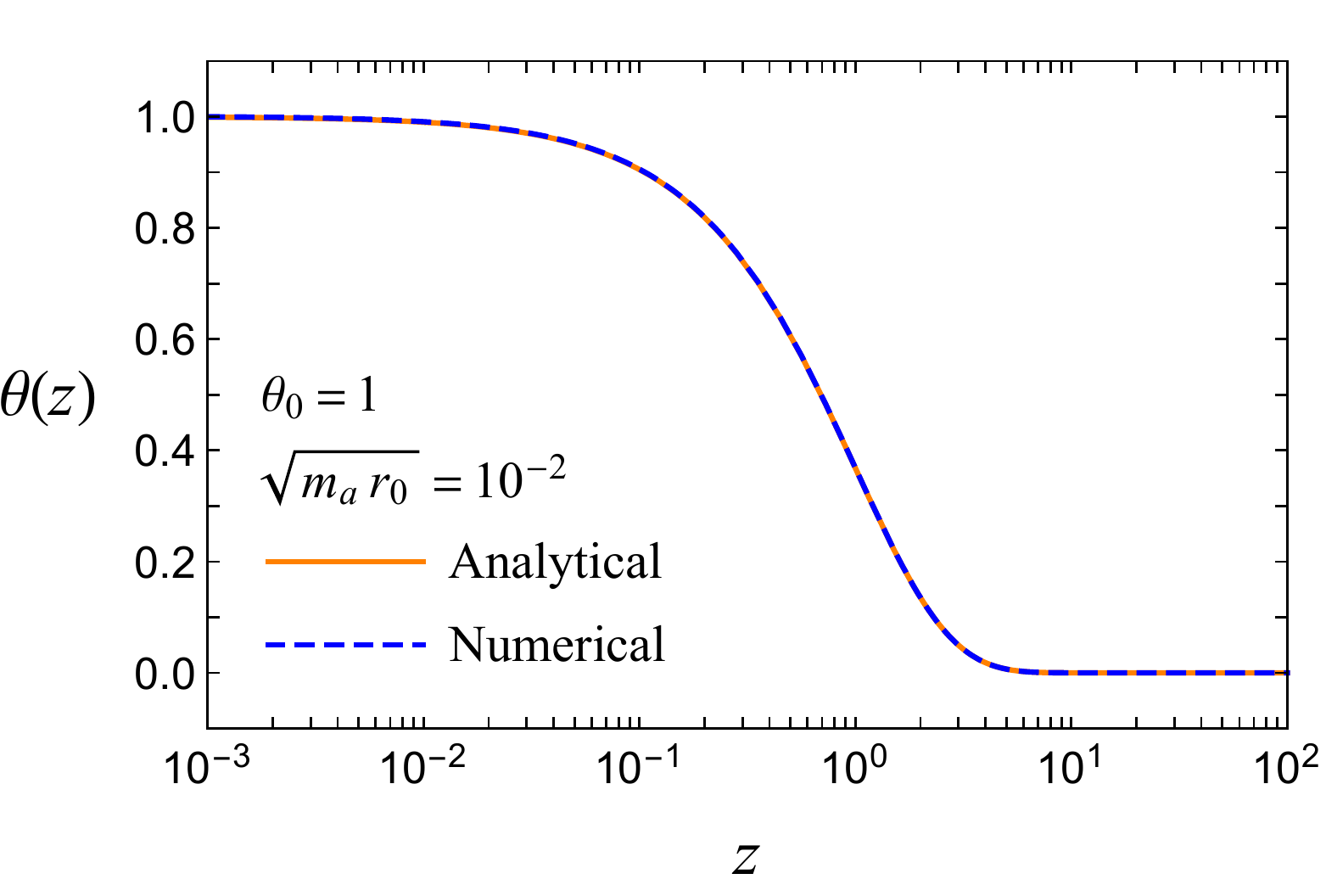}
\vspace{-0.3cm}
\caption{The configuration of the axion field  around the monopole as a function of $z = r_0/r$
for $\theta_0 = 1$ and $\sqrt{m_a r_0} = 10^{-2}$.
The monopole is located at $z \to \infty$ (i.e.\,$r =0$), and the axion field takes the vacuum value $\theta_0$ 
at $z \to 0$ \,(i.e.\,$r \to \infty $). }
\label{fig:theta_vs_z}
\end{figure}

What we want to know is the field configuration that minimizes the Hamiltonian. To this end, it is useful to change the radial coordinate $r$ to a dimensionless variable $z = r_0/r$.
The equation to solve is
\begin{eqnarray}\label{EOM}
\frac{d^2\theta(z)}{dz^2} 
\,=\, 
\theta(z) + \frac{m_a^2 r_0^2}{z^4}
\big(\theta(z)-\theta_0\big)
~.
\end{eqnarray} 
In order for the energy density of the axion-hidden monopole system to be finite, we also impose $\theta(z\to\infty) = 0$ and $\theta(z\to 0) = \theta_0$ as the boundary conditions of Eq.\,\eqref{EOM}. Unfortunately, one cannot solve this differential equation analytically. Instead, let us solve this equation with its asymptotic forms as
\begin{eqnarray}
\frac{d^2\theta}{dz^2} 
\,\simeq\, 
\begin{cases}
\displaystyle\,
\theta 
& \text{for} ~~ z > \sqrt{m_a r_0}
\\[0.5cm]
\displaystyle\,
\frac{m^2_a r^2_0}{z^4}
(\theta-\theta_0)
& \text{for} ~~ z < \sqrt{m_a r_0}
\end{cases} 
~.
\end{eqnarray}
This has the analytical solutions given by
\begin{eqnarray}\label{thetasol}
\theta(z) 
\,\simeq\, 
\begin{cases}
\displaystyle\,
\theta_>(z) \equiv\,
\theta_0 
\bigg(
\frac{1+\sqrt{m_a r_0}}{1+2\sqrt{m_a r_0}}
\bigg)
e^{-z+\sqrt{m_a r_0}}
&\text{for} ~~ z > \sqrt{m_a r_0}
\\[0.6cm]
\displaystyle\,
\theta_<(z) \equiv\,
\theta_0 
\bigg(
1-\frac{z}{1+2\sqrt{m_a r_0}}\,e^{-m_a r_0/z+\sqrt{m_a r_0}}
\bigg)
&\text{for} ~~ z < \sqrt{m_a r_0}
\end{cases} 
~.
\end{eqnarray}
Notice that this solution is smooth at $z = \sqrt{m_a r_0}$, and we have checked that the asymptotic solutions given by \eqref{thetasol} agree with the numerical solution of \eqref{EOM} to very high precision. See Fig.~\ref{fig:theta_vs_z}, where we make a comparison between the analytic and numerical solutions for  $\theta_0 = 1$ and $\sqrt{m_a r_0} = 10^{-2}$. One can see that the two lines agree very well. The monopole is located at $z \to \infty$ (i.e.\,$r =0$), and the axion field takes the vacuum value $\theta_0$ at $z \to 0$ \,(i.e.\,$r \to \infty $). The axion field value starts to decrease around $z \simeq 1$\,(i.e.\,$r \simeq r_0$), and this is due to the fact that a nonzero axion field costs large energy stored in the hidden electric field near the origin because of $E_{\rm H}(r)^2 \propto \theta(r)^2/r^4$.

So far we have not specified the relationship between $m_a$ and $1/r_0$. For our purpose, we are interested in relatively light axion masses between $10$\,MeV and $10$\,GeV (the mass region of scalar particles searched by the beam-dump experiments), while the decay constant $f^{}_{\rm H}$ is constrained to be much larger than $m_a$ for this mass range. In the following calculation, we, therefore, assume $m_a r_0 \ll 1$, for which the axion mass is relevant only in the tail of the axion field configuration around the monopole. As we shall see, those outer parts contribute most to the scattering cross section, because the typical momentum transfer between the monopole and the nucleon is  smaller than $m_a$. It is also clear that the precise size of the monopole core radius does not affect the calculation of the cross section.

\subsection{Spin-dependent monopole-nucleon scattering cross section}
With the axion field profile \eqref{thetasol}, we can now calculate the amplitude of the monopole-nucleon scattering by introducing the following axion-nucleon interaction
\begin{eqnarray}\label{aN}
H_{a-{\rm N}} 
\,=\,
\frac{C_{\rm N}m_{\rm N}}{f_a} 
\mathop{\mathlarger{\int}} \hspace{-0.1cm} d^3 x \,
\Big[
a(x)
\overline{\psi}_{\rm N}(x) i\gamma^5 \psi^{}_{\rm N}(x)
\Big]
~,
\end{eqnarray}
where N presents the nucleon (proton $p$ or neutron $n$), $C_{\rm N}$ is a constant of order unity, $m_{\rm N} \simeq 1\,\text{GeV}$ is the nucleon mass, and $f_a$ is a mass scale characterizing the strength of the axion-nucleon coupling. 
In the following calculation, we will assume $f_a = f^{}_{\rm H}$ for simplicity, but $f_a$ can be much larger or smaller than $f_{\rm H}$ in a slightly contrived setup. The values of $C_{p}$ and $C_{n}$ are actually model-dependent. In our numerical study, we choose $|C_p| \simeq 0.4$ and $|C_n| \simeq 0.05$, which are typical values in the DFSZ axion model~\cite{diCortona:2015ldu}.

In our set-up, the monopole is much heavier than nucleons, and so, we can treat $a(x)$ as a classical scalar field around the monopole while we treat $\psi^{}_{\rm N}(x)$ as the usual quantized Dirac field in the same way as the Rutherford scattering. Then, the amplitude of the monopole-nucleon scattering mediated by the axion at the leading order  is written as~\cite{Peskin:1995ev}
\begin{eqnarray}\label{MNMN}
i{\cal M}_{{\rm M}+{\rm N}\to{\rm M}+{\rm N}} 
\,=\, 
C_{\rm N}m_{\rm N} \, \widetilde{\theta}(\boldsymbol{q})\, 
\overline{u}_{\rm N}(p') \gamma^5 u_{\rm N}(p) ~,
\end{eqnarray}
where $u_{\rm N}$ is the Dirac spinor, $\boldsymbol{q}=\boldsymbol{p}'-\boldsymbol{p}$ is the transferred 3-momentum,
and $\widetilde{\theta}(\boldsymbol{q})$ is the 3-dimensional Fourier transform of $\theta(x)$ given by
\begin{eqnarray}\label{thetaq}
\widetilde{\theta}(\boldsymbol{q})
\;\equiv\;
\mathop{\mathlarger{\int}} \hspace{-0.1cm} 
d^3 x\, \theta(\mathbf{x})\, e^{-i\boldsymbol{q} \cdot \mathbf{x}}
\,=\,
\frac{4\pi r_0^2}{|\boldsymbol{q}|} 
\mathop{\mathlarger{\int}_0^{\,r_0/r_{\rm c}}} \hspace{-0.1cm} dz\,
\frac{\theta(z)}{z^3} \sin\bigg(\frac{|\boldsymbol{q}|r_0}{z}\bigg)
~.
\end{eqnarray}
Here we have introduced the cut off in the integration corresponding to the core radius because the axion configuration derived before is valid only for $r \gtrsim r_{\rm c}$. Plugging Eq.\,\eqref{thetasol} into Eq.\,\eqref{thetaq}, we have
\begin{eqnarray}\label{thetacore1}
\widetilde{\theta}(\boldsymbol{q})
\Eq
\frac{4\pi r_0^2}{|\boldsymbol{q}|} 
\Bigg[
\mathop{\mathlarger{\int}_0^{\sqrt{m_a r_0}}}  dz\,
\frac{\theta_<(z)}{z^3}\sin\bigg(\frac{|\boldsymbol{q}|r_0}{z}\bigg)
+
\mathop{\mathlarger{\int}_{\sqrt{m_a r_0}}^{\,r_0/r_{\rm c}}}  dz\,
\frac{\theta_>(z)}{z^3}\sin\bigg(\frac{|\boldsymbol{q}|r_0}{z}\bigg)
\Bigg] 
\end{eqnarray}
for $\sqrt{m_a r_0} < r_0/r_{\rm c}$, and
\begin{eqnarray}\label{thetacore2}
\widetilde{\theta}(\boldsymbol{q})
\Eq
\frac{4\pi r_0^2}{|\boldsymbol{q}|} 
\mathop{\mathlarger{\int}_0^{\,r_0/r_{\rm c}}}  dz\,
\frac{\theta_<(z)}{z^3}\sin\bigg(\frac{|\boldsymbol{q}|r_0}{z}\bigg)
~
\end{eqnarray}
for $\sqrt{m_a r_0} > r_0/r_{\rm c}$.
The integration of \eqref{thetacore1} and \eqref{thetacore2} involves an IR divergence at $z \to 0$, which can be removed by inserting the regulator, $\lim_{\delta \to0}e^{-\delta /z}$. Then, both integrations give approximately the same result,
\begin{eqnarray}\label{thetaqresult}
\widetilde{\theta}(\boldsymbol{q})
\,\simeq\,
-4\pi \theta_0 
\bigg(
\frac{r_0}{m_a^2} +
\frac{r_{\rm c}^3}{3}  
\bigg)
+
{\cal O}\big(\epsilon_q^2\big)
\end{eqnarray}
for $\sqrt{m_a r_0} \ll 1$.\footnote{If $m_a \lesssim 10^{-3}\,\text{GeV}$, then one has to do the integrations numerically,
because $\epsilon_q$ is no longer smaller than unity. For the interesting range of the decay constant, such a light axion is tightly constrained by astrophysics\,\cite{Tanabashi:2018oca}, and we do not consider it in this paper.} 
Here we have defined $\epsilon_q = |\boldsymbol{q}|/m_a \simeq m_{\rm N}\upsilon_{\rm DM}/m_a \ll 1$, where $\upsilon_{\rm DM} \simeq 10^{-3}$ is the typical velocity of DM with respect to the reference frame of the nucleon at the solar radius 
in our Galaxy.  Also, we have numerically checked that Eq.\,\eqref{thetaqresult} gives a good approximation to Eqs.\,\eqref{thetacore1} and \eqref{thetacore2} as long as $\sqrt{m_a r_0} \ll 1$ no matter the size of $r_0/r_{\rm c}$. This is because
the DM-nucleon scattering is a low-energy process and therefore the integral is dominated by small $z$ (i.e. large $r$).

For our benchmark point, the second term in the bracket of \eqref{thetaqresult} is negligible since we consider very heavy
monopole mass, namely the effect of the hidden monopole core can be neglected. Then, with Eqs.\,\eqref{MNMN} and \eqref{thetaqresult}, the resulting differential spin-dependent cross-section of the hidden monopole DM elastic scattering off a nucleon is evaluated as~\cite{Peskin:1995ev}
\begin{eqnarray}\label{sigmaMNMN}
\frac{d\sigma_{{\rm M}+{\rm N}\to{\rm M}+{\rm N}}}{d\Omega}
\,\simeq\,
\frac{\alpha^{}_{\rm H}\theta_0^2 C_{\rm N}^2}{16\pi^3} 
\frac{m^2_{\rm N}}{ m_a^4 f^2_a} 
|\boldsymbol{q}|^2   
~.
\end{eqnarray}
We will use this result to constraint the axion decay constant by direct DM detection experiments in the next subsection.

Before ending this subsection, let us briefly discuss the hidden vector boson DM in our model setup since it can also interact with  nucleus through the axion exchange. The elastic scattering of the hidden vector boson DM and nucleon can be described by the  following dimensional-7 effective operator as
\begin{eqnarray}
{\cal O}^{(7)}_{{\rm N}W}
\,=\,
\frac{\alpha^{}_{\rm H}}{8\pi}
\frac{m_{\rm N}}{m_a^2 f_a^2} 
\overline{\psi}_{\rm N} i\gamma_5 \psi^{}_{\rm N} 
W_{\rm H}^{\pm\mu\nu} \widetilde{W}^{\mp}_{{\rm H}\mu\nu} 
~,
\end{eqnarray}
and the resulting differential cross-section of the hidden gauge boson scattering off a
nucleon is computed as
\begin{eqnarray}\label{sigmaWNWN}
\frac{d\sigma_{W + {\rm N} \to W+ {\rm N}}}{d\Omega}
\,\simeq\, 
\frac{\alpha^2_{\rm H}}{1536\pi^4}
\frac{m_{\rm N}^2}{m_a^4 f_a^4} |\boldsymbol{q}|^4 ~.
\end{eqnarray}
Comparing Eq.\,\eqref{sigmaWNWN} with Eq.\,\eqref{sigmaMNMN}, we find
\begin{eqnarray}
\frac
{d\sigma_{W + {\rm N} \to W+ {\rm N}}/d\Omega}
{d\sigma_{{\rm M} + {\rm N} \to {\rm M}+ {\rm N}}/d\Omega}
\,\simeq\, 
\frac{\alpha^{}_{\rm H}}{96\pi \theta_0^2C_{\rm N}^2}\frac{|\boldsymbol{q}|^2}{f_a^2} 
\,\ll\, 1 ~,
\end{eqnarray}
thereby the hidden gauge boson-nucleon scattering is negligible, which motivates us to focus on the hidden monopole-nucleon scattering via the axion portal.

Lastly, with the results of \eqref{sigmaMNMN} and \eqref{sigmaWNWN} one may wonder why the hidden monopole-nucleon scattering cross-section is suppressed only by the decay constant squared rather than fourth power of it. If one goes back to the calculation, one can see that the peculiar dependence on $f_a$ comes from the axion configuration around the monopole, and that the dominant contribution to the scattering cross section comes from  $r \sim \sqrt{r_0/m_a}$, where the axion mass becomes relevant.

\subsection{Limits and forecasts of direct DM search experiments}
Now let us estimate the present limits on the predicted scattering cross section \eqref{sigmaMNMN} from the direct DM search experiments. To this end, one can consider the following dimensional-6 pseudo-scalar operator given in Ref.\,\cite{Dienes:2013xya} since it gives the same dependence of the scattering cross section on the transferred momentum $\boldsymbol{q}$,
\begin{eqnarray}
\label{dim6}
{\cal O}^{(6)}_{\chi{\rm N}}
\,=\,
\frac{g^{}_{\chi \text{N}}}{\Lambda^2}
\big(\,\overline{\psi}_\chi \psi^{}_\chi \big)
\big(\,\overline{\psi}_{\rm N} i\gamma^5 \psi^{}_{\rm N}\big) 
~,
\end{eqnarray}
where $\chi$ is a Dirac fermionic DM with $g^{}_{\chi \text{N}}$ being the coupling strength to the nucleon, and $\Lambda$ is the cut-off scale of the theory.\footnote{See Refs.~\cite{Abe:2018emu,Ertas:2019dew,Li:2019fnn} for the loop corrections in a pseudoscalar mediator DM model.} The tree-level matrix element of the $\chi$-${\rm N}$ scattering in the non-relativistic limit is given by
\begin{eqnarray}
{\cal M}_{\chi + {\rm N} \to \chi + {\rm N}}
\Eq
\frac{2 g^{}_{\chi \text{N}} m_{\rm DM}}{\Lambda^2} 
\big[ \xi^{\dagger}_{\rm \chi}(s'_\chi) \xi^{}_{\rm \chi}(s_\chi) \big]
\big[ \xi^{\dagger}_{\rm N}(s'_{\rm N}) (\boldsymbol{q} \cdot \boldsymbol{\sigma}) \xi^{}_{\rm N}(s_{\rm N}) \big] ~,
\end{eqnarray}
where $m_{\rm DM}$ is the mass of $\chi$, $\xi(s)$ is the two-component spinor with $s$ being the state of spin, and $\boldsymbol{\sigma}=(\sigma_1,\sigma_2,\sigma_3)$ is the Pauli spin matrix vector. The spin-averaged differential cross-section is then estimated as
\begin{eqnarray}\label{sigmapsiN}
\frac{d\sigma_{\chi + {\rm N} \to \chi + {\rm N}}}{d\Omega}
\,\simeq\, 
\frac{\overline{\big|{\cal M}_{\chi + {\rm N} \to \chi + {\rm N}}\big|^2} }{64\pi^2(m_{\rm DM}+ m_{\rm N})^2}
\,\simeq\, 
\frac{g^2_{\chi \text{N}}}{16\pi^2\Lambda^4} |\boldsymbol{q}|^2 ~,
\end{eqnarray}
where we have assumed that $m_{\rm DM} \gg m_{\rm N}$ in the last equality. Thus, one can see that this has the same momentum dependence as \eqref{sigmaMNMN}.  Note also that the operator (\ref{dim6}) violates CP corresponding to the fact that the axion configuration around the monopole breaks CP.

In our study, we will focus on the case of proton since the COUPP and PICO experiments are sensitive to the
DM  scattering cross section with a proton. Making a comparison of \eqref{sigmaMNMN} with \eqref{sigmapsiN} and referring the results analyzed in Ref.\,\cite{Dienes:2013xya}, we present the constraints on the axion decay constant from the direct DM search experiments as a function of the axion mass with different choices of $\alpha^{}_{\rm H}$ and $m_{\rm M}$ in Fig.~\ref{fig:fainv_vs_ma}, where the amount of DM interacting with the nucleon is taken into account as explained in the last subsection. In these plots, the red and green shaded regions are excluded by the current experiments COUPP-4 (4.0 kg CF$_3$I)\,\cite{Behnke:2012ys} and PICO-60 (52 kg C$_3$F$_8$)\,\cite{Amole:2019fdf}, respectively, and the blue dashed line is the future prospect by PICO-500 (C$_3$F$_8$)\,\cite{PICO500}.
\begin{figure}[t]
\centering
\includegraphics[scale=0.5]{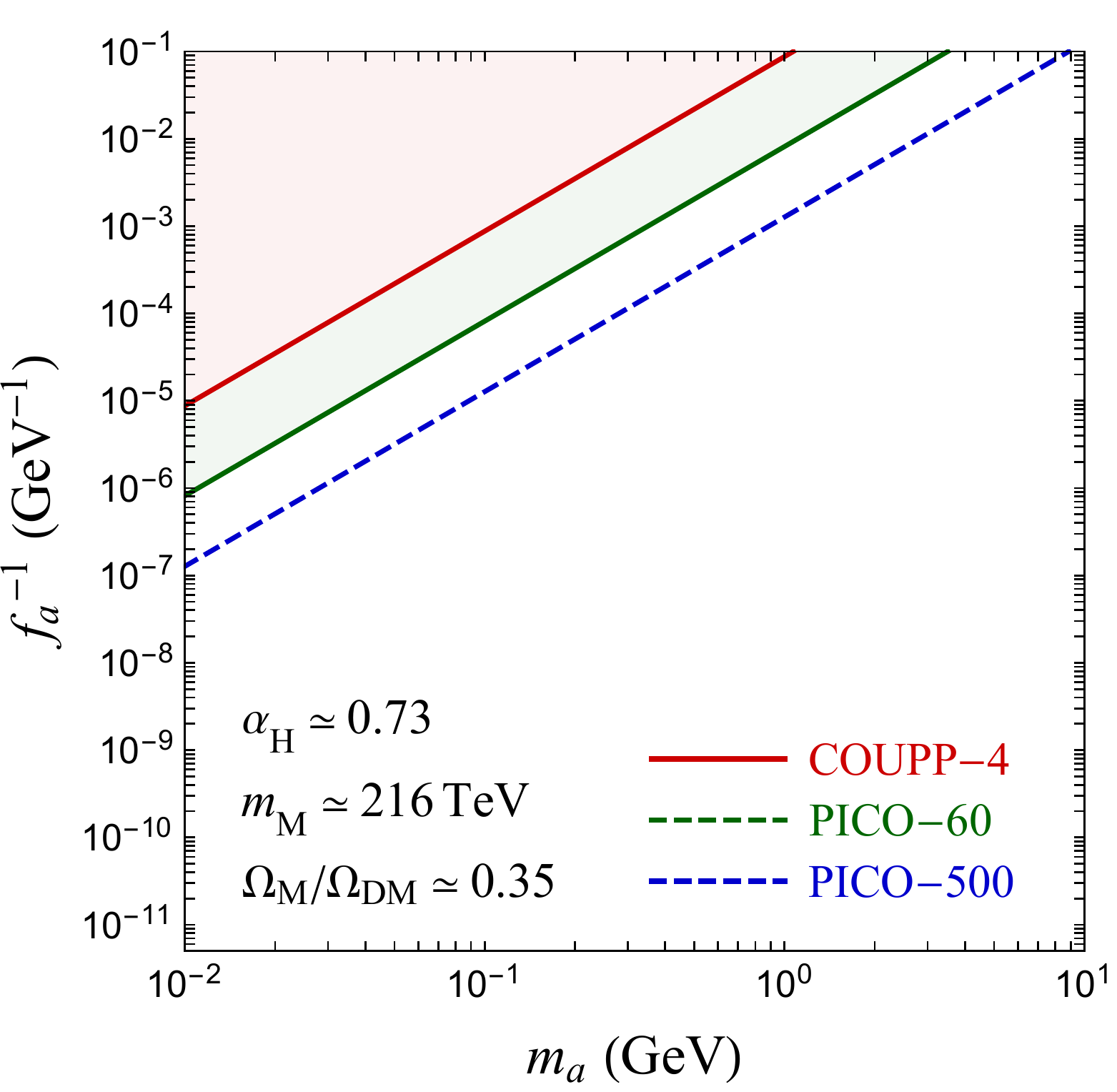}
\\[0.3cm]
\includegraphics[scale=0.5]{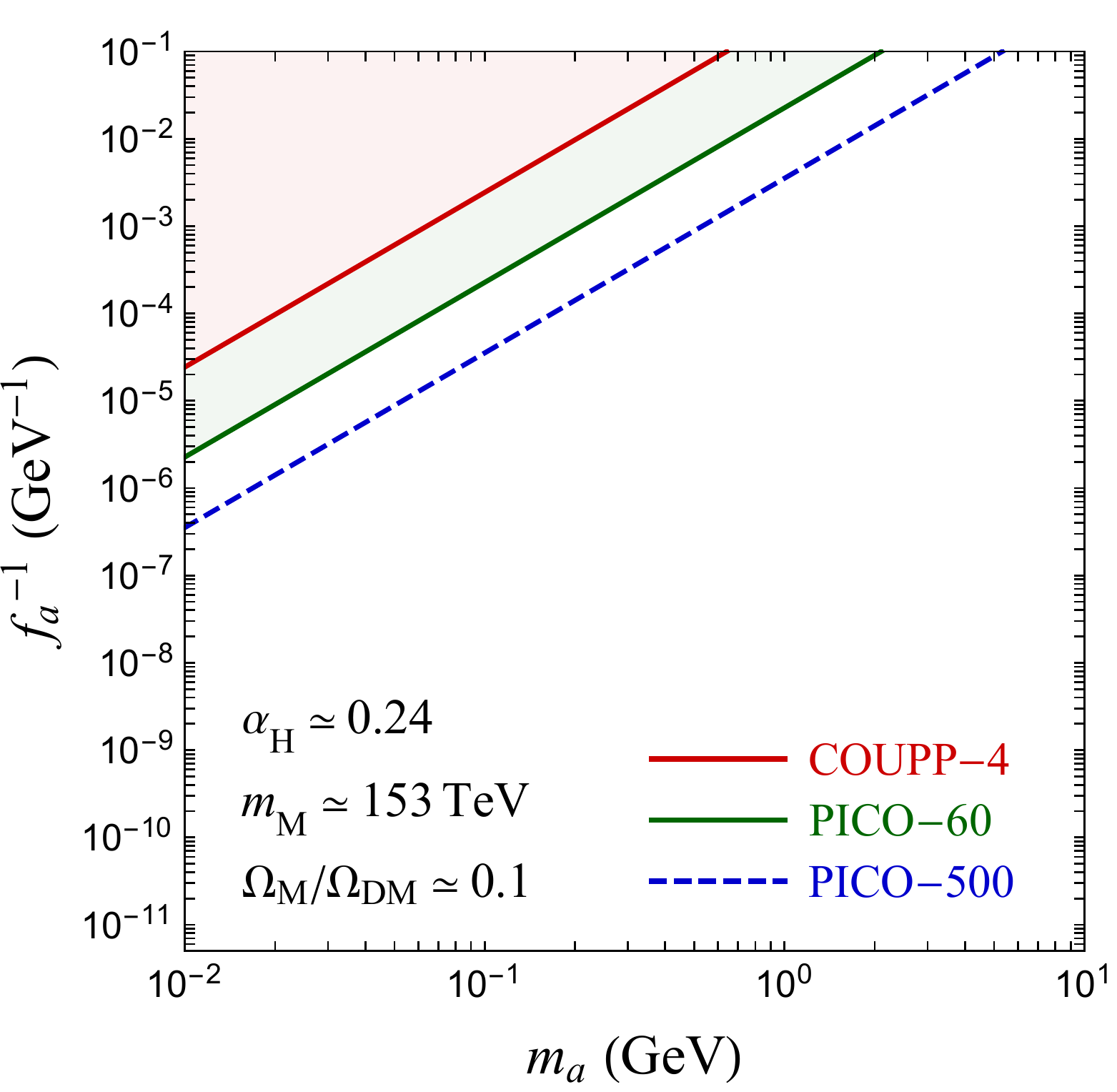}
\hspace{0.1cm}
\includegraphics[scale=0.5]{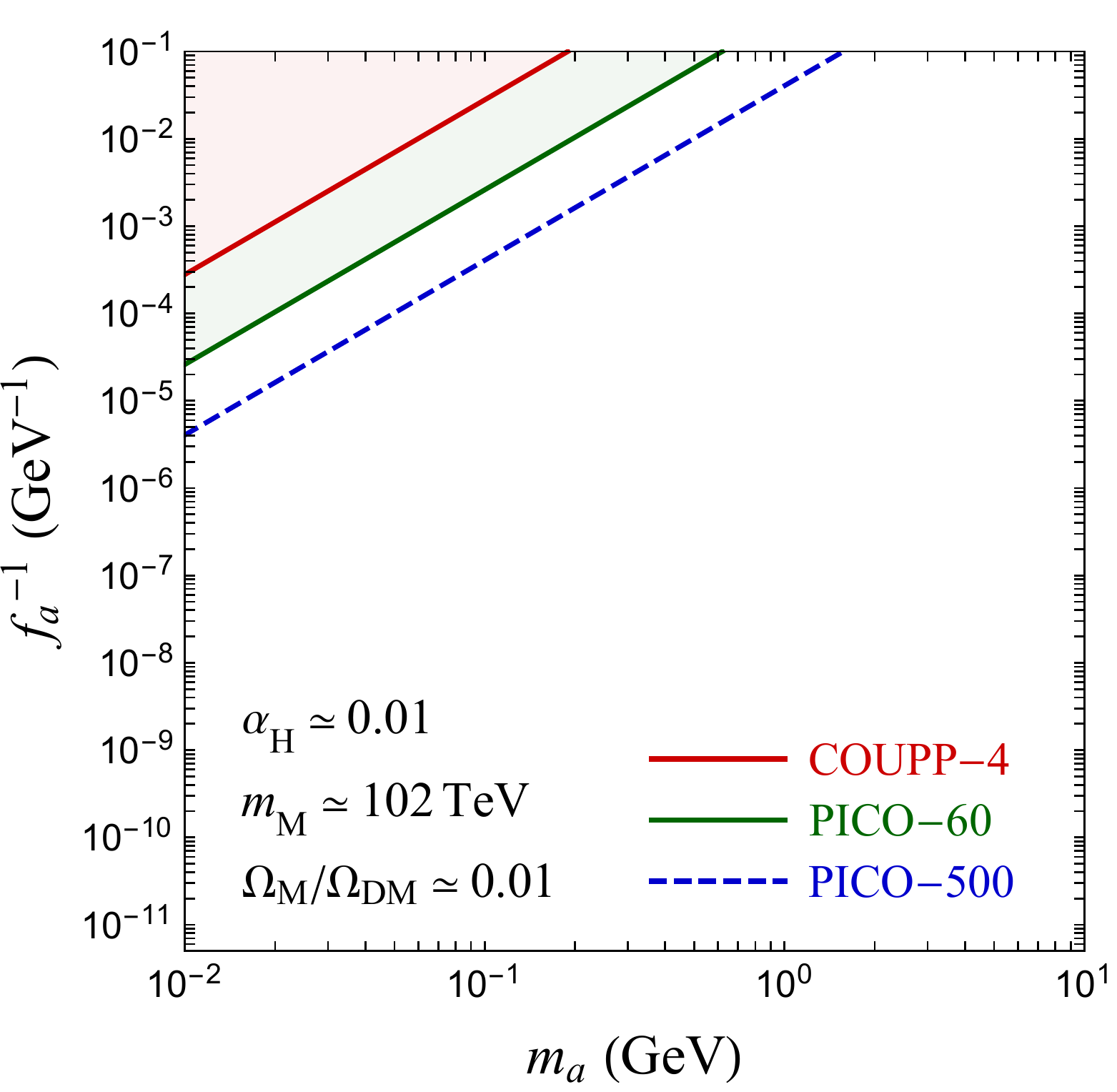}
\vspace{-0.3cm}
\caption{The lower bound of the axion decay constant as a function of the axion mass by imposing the constraints from direct DM search experiments with different choices of $\alpha^{}_{\rm H}$ and $m_{\rm M}$, where the red and green shaded regions are ruled out by the current experiments COUPP-4 and PICO-60, respectively, and the blue dashed line is the future sensitivity by PICO-500. The percentage of DM interacting with the nucleon is considered, which is indicated by the ratio $\Omega_{\rm M}/\Omega_{\rm DM}$ in these plots. Here we have considered the case of proton and fixed $\theta_0 =1$.}
\label{fig:fainv_vs_ma_DD}
\end{figure}

\section{Implications for axion search experiments}
\label{sec:4}
In this section, we study the current limits and future prospects for the search of axions. In our setup \eqref{axion_portal}, the axion is coupled not only to the SM particles but also to the hidden photons. The axion can decay into a pair of the hidden photons with the  rate given by
\begin{eqnarray}\label{A1}
\Gamma(a \to \gamma^{}_{\rm H} \gamma^{}_{\rm H}) \,=\, 
\frac{\alpha^2_{\rm H} m_a^3}{256\pi^3 f_{\rm H}^2} 
~.
\end{eqnarray}
This new decay mode changes the bounds on the axion decay constant from the beam-dump experiments (CHARM, SHiP, etc.) and some $B$-meson rare decays. We briefly write down the relevant formulas and constraints in the following. For simplicity, we will set $f_{\rm H} = f_a$. 

The beam-dump of CHARM (SHiP) experiment imposes that the number of events $N_{\rm det}$ in the detector region is~\cite{Dolan:2014ska,Clarke:2013aya,Alekhin:2015byh}
\begin{eqnarray}\label{A2}
N_{\rm det} \approx
N_a 
\exp\bigg({-}\frac{\ell}{\gamma_a \beta_a c\,\tau_a}  \bigg)
\bigg[
1 - \exp\bigg({-}\frac{\Delta\ell}{\gamma_a \beta_a c\,\tau_a}\bigg)
\bigg]
\sum_{X=e,\,\mu,\gamma} 
{\cal B}(a \to X\bar{X})
\,<\, 2.3\, (3) ~,
\end{eqnarray}
where $N_a$ is the number of the axion produced in the solid angle of the detector with length $\Delta\ell =35\,(55)\,\text{m}$, $\gamma_a = (1-\beta^2_a)^{-1/2} \approx 10\,(25)\,\text{GeV}/m_a $\,\cite{Bezrukov:2009yw,Alekhin:2015byh}, $c$ is the speed of light, and $\ell = 480\,(70)\,\text{m}$ is the distance between the detector and the beam-dump. The lifetime of the axion $\tau_a$ reads
\begin{eqnarray}\label{A3}
\tau_a 
\,=\, \frac{1}{\Gamma_a} 
\,=\, \frac{1}{\Gamma(a \to\gamma^{}_{\rm H} \gamma^{}_{\rm H}) + \Gamma(a \to {\rm vis})} ~,
\end{eqnarray}
where $\Gamma_a$ is the total width of the axion, and $\Gamma(a \to\gamma^{}_{\rm H} \gamma^{}_{\rm H})$ and $\Gamma(a \to {\rm vis})$ are the rates of the axion decaying into the hidden photons and visible particles ($e^{\pm}, \mu^{\pm}, \gamma,\cdots$), respectively.

On the other hand, the upper bound of the branching ratio of $B^0$ decaying into $K_S^0$ plus missing energy is given by~\cite{Dolan:2014ska,Ammar:2001gi}
\begin{eqnarray}\label{A4}
{\cal B}\big(B\to K+{\rm inv})\,=\,
{\cal B}\big(B\to Ka\big)P_{\rm esc}(\boldsymbol{p}_a) 
\,<\,  5.3 \times 10^{-5}
~,
\end{eqnarray}
where $P_{\rm esc}(\boldsymbol{p}_a)$ is the probability of the axion with momentum $\boldsymbol{p}_a$ 
to escape from the detector 
\begin{eqnarray}\label{A5}
P_{\rm esc}(\boldsymbol{p}_a) 
\,=\, 
{\cal B}(a \to\gamma^{}_{\rm H} \gamma^{}_{\rm H})
+
{\cal B}(a \to {\rm vis})
\exp\bigg({-}\frac{\ell_{\rm max} m_a \Gamma_a}{|\boldsymbol{p}_a|}\bigg)
\end{eqnarray}
with ${\cal B}(a \to\gamma^{}_{\rm H} \gamma^{}_{\rm H}) = \Gamma(a \to\gamma^{}_{\rm H} \gamma^{}_{\rm H})/\Gamma_a$ and ${\cal B}(a \to {\rm vis}) = \Gamma(a \to {\rm vis})/\Gamma_a$ the branching ratios of the axion decaying into the hidden photons and visible particles, respectively. In our numerical study, we take $\ell_{\rm max} = 4\,{\rm m}$ as required in Ref.\,\cite{Dolan:2014ska}.

\begin{figure}[t]
\centering
\includegraphics[scale=0.5]{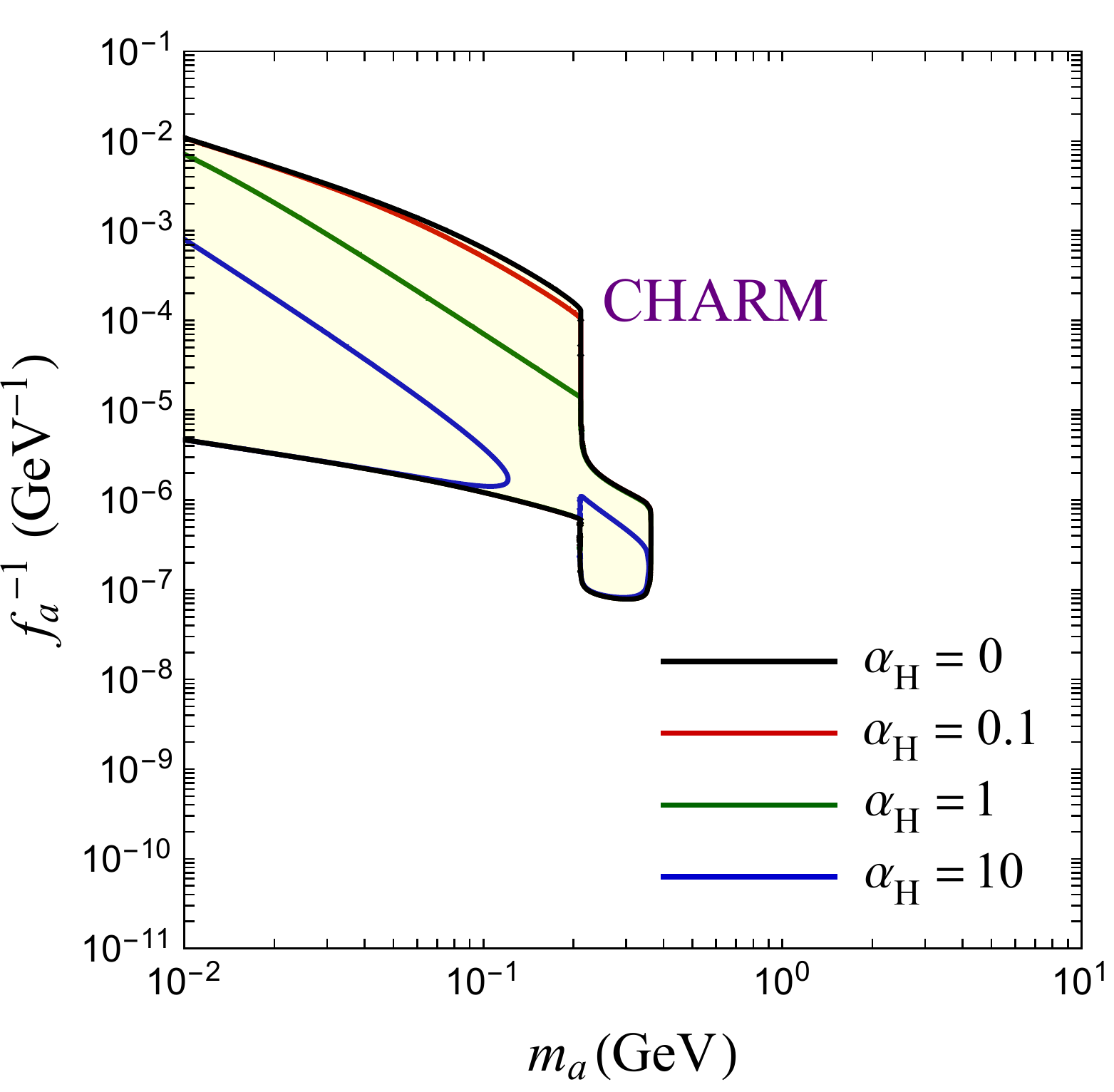}
\hspace{0.1cm}
\includegraphics[scale=0.5]{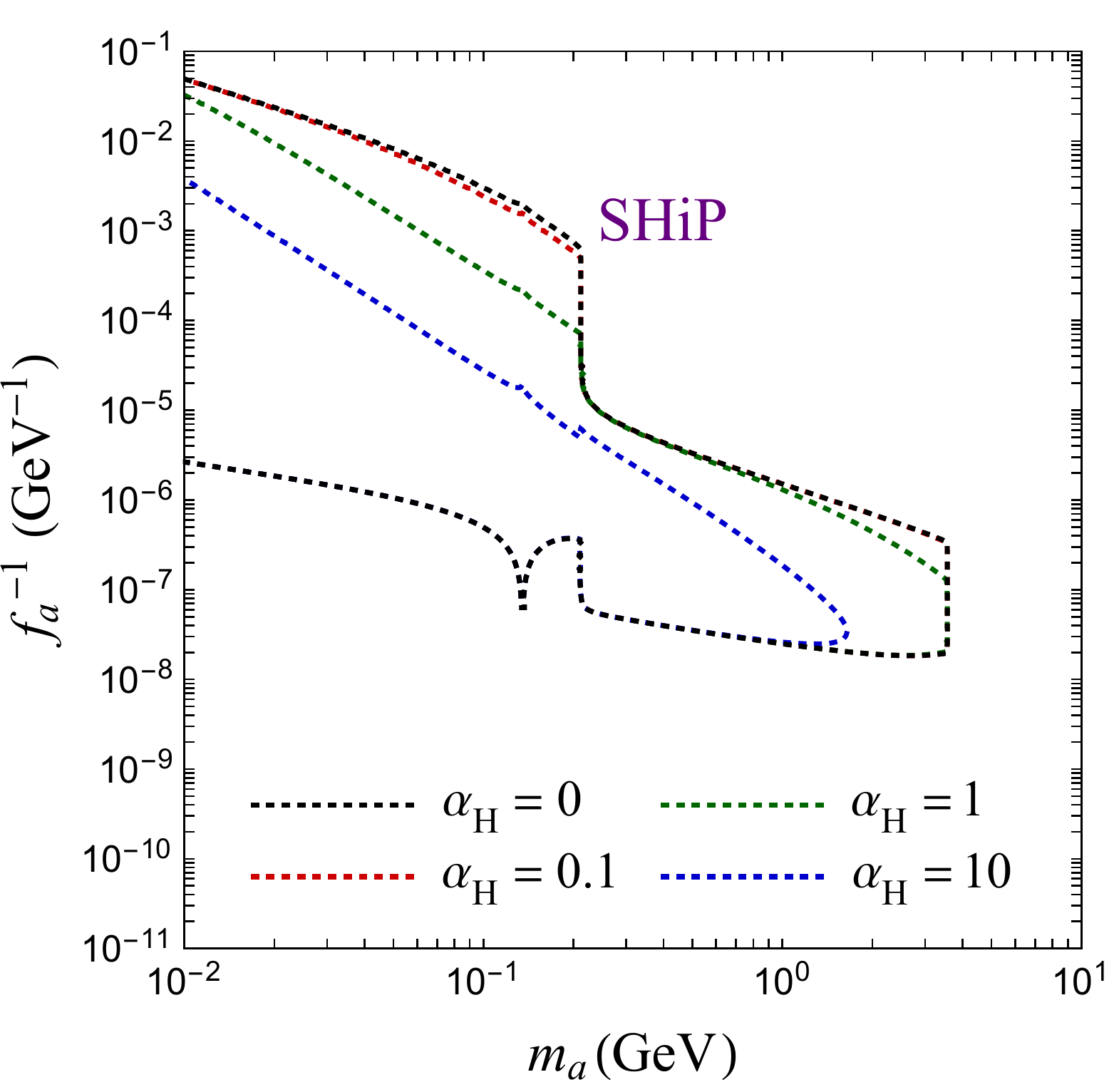}
\\[0.3cm]
\includegraphics[scale=0.5]{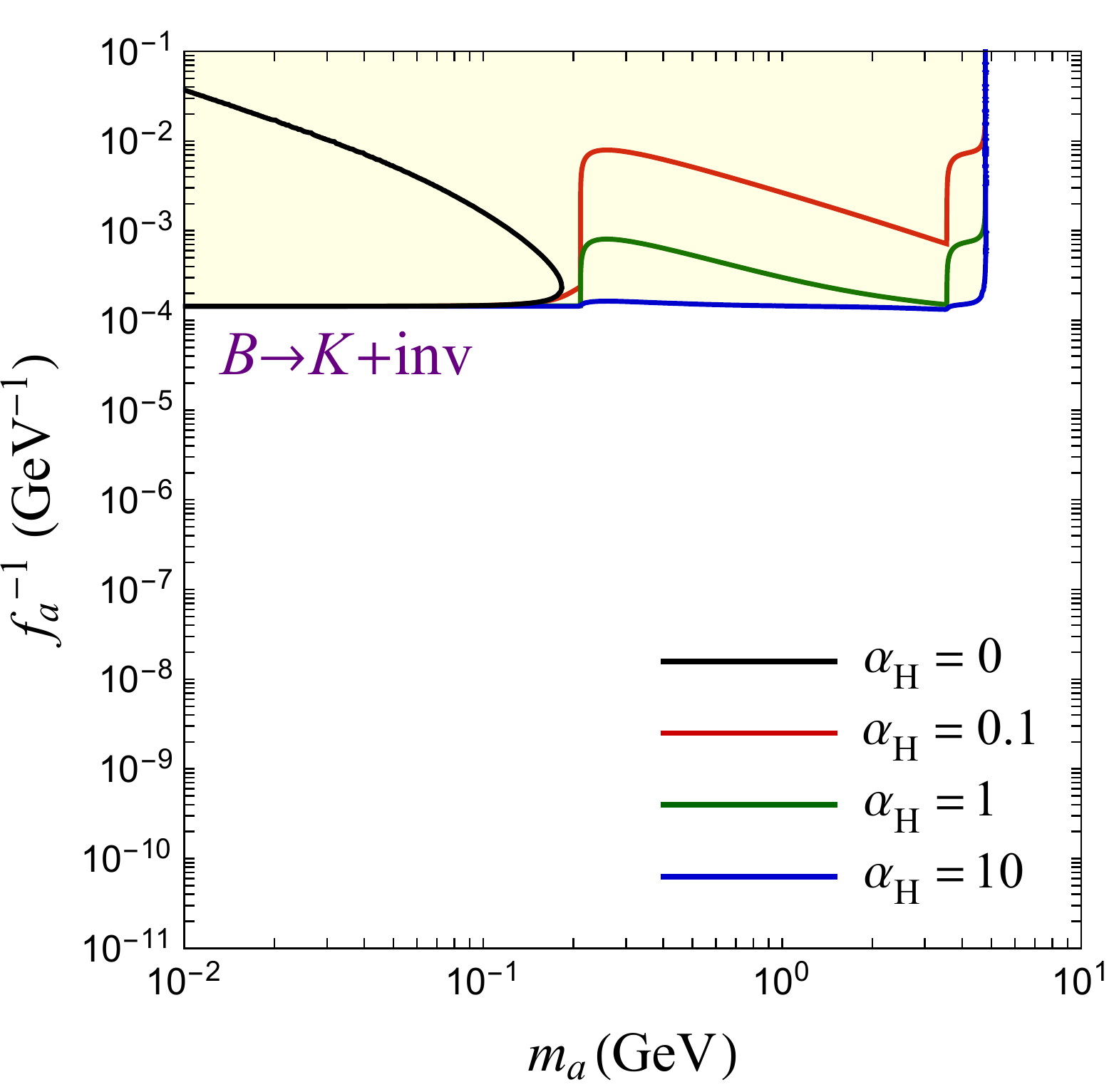}
\caption{The effects of the decay channel $a\to \gamma^{}_{\rm H} \gamma^{}_{\rm H}$ on the CHARM experiment (upper left-panel), the SHiP experiment (upper right-panel), and $B^0$ decaying into $K^0_S$ plus missing energy (bottom) with different choices of $\alpha^{}_{\rm H}$, assuming the Yukawa-like coupling for the axion, where the yellow shaded regions are excluded by current experiments.
We set $f_{\rm H} = f_a$, for simplicity. 
} 
\label{fig:fainv_vs_ma_CHARM_Bmeson}
\end{figure}

Applying Eq.\,\eqref{A1} to \eqref{A5}, we present in Fig.~\ref{fig:fainv_vs_ma_CHARM_Bmeson}, how the decay channel $a\to \gamma^{}_{\rm H}\gamma^{}_{\rm H}$ affects the limits from the CHARM experiment, the SHiP experiment, and $B^0$ decaying into $K^0_S$ with missing energy. The constraints from $B \to K\mu^+\mu^-$ and $B \to \mu^+\mu^-$ are not shown here since we have checked that the effects of them are weaker than the others. Notice that we assume the Yukawa-like coupling for the axion in our analysis,
\begin{eqnarray}
{\cal L}_Y \,=\,
\sum_{f} 
\frac{m_f}{f_a}\,a \overline{f} i \gamma_5 f 
~,
\end{eqnarray}
where the couplings between the axion and the SM fermions are proportional to the SM Yukawa couplings\,\cite{Dolan:2014ska}. This interaction may be induced in a set-up like the DFSZ axion model\,\cite{Dine:1981rt,Zhitnitsky:1980tq}.

In the upper plots of Fig.~\ref{fig:fainv_vs_ma_CHARM_Bmeson}, the parameter region of $f_a^{-1}$ above the color lines corresponds to the case where the axion decays before arriving the detector. Due to the hidden photon decay mode of the axion, the upper parts of these color lines are shrunk as one has to reduce the coupling strength $(\propto f_a^{-1})$ between the axion and the SM particles in order to maintain the same predicted event numbers. On the other hand, the parameter region of $f_a^{-1}$ below the color lines corresponds to the case where the axion decays after leaving the detector. In this region, the lifetime of the axion is very long such that the predicted event numbers is $N_a \Delta \ell \sum_{X = e,\mu,\gamma} \Gamma(a \to X\bar{X})/(\gamma_a \beta_a c)$, which is independent of $\alpha^{}_{\rm H}$. This explains why the lower parts of the color lines are insensitive to $\alpha^{}_{\rm H}$. In the bottom plot of Fig.~\ref{fig:fainv_vs_ma_CHARM_Bmeson}, the excluding region of $f_a^{-1}$ is enlarged since we identify with the axion decays into the hidden photons within the detector as an  escaped event, then ${\cal B}(a\to\gamma_{\rm H}\gamma_{\rm H})$ gives a non-negligible contribution in Eq.\,\eqref{A5} even it is a sub-dominant decay mode at $m_a > 2m_{\mu}$.

\begin{figure}[t]
\centering
\includegraphics[scale=0.6]{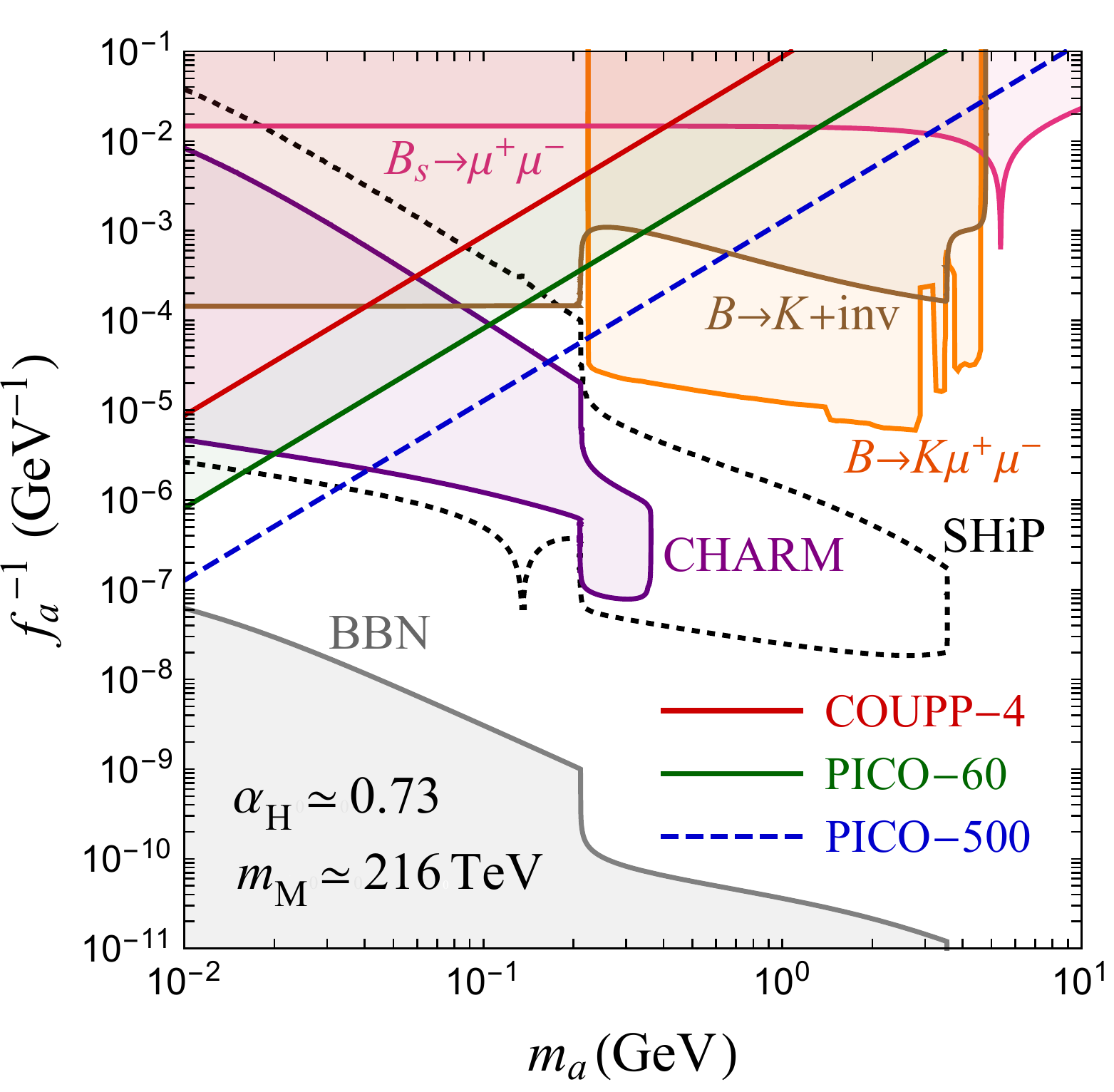}
\caption{The axion decay constant versus the axion mass, where the shaded regions are excluded by the current experiments such as COUPP-4, PICO-60, CHARM, and some $B$-meson rare decays and by the cosmological constraints like BBN. On the other hand, the dashed line and the dotted curve indicate the future sensitivities by PICO-500 and SHiP experiments, respectively. In addition, we have fixed $\theta_0 =1$ and $f^{}_{\rm H} = f_a$. For the BBN constraint, we simply require that the lifetime of the axion should be less than one second, a time when the BBN starts to occur after the big bang.} 
\label{fig:fainv_vs_ma}
\end{figure}

Combining the results from Fig.\,\ref{fig:fainv_vs_ma_DD} and Fig.\,\ref{fig:fainv_vs_ma_CHARM_Bmeson} with our benchmark point, we present the upshot of this paper in Fig.\,\ref{fig:fainv_vs_ma}, where the constraints from $B$-meson decays $B \to K\mu^+\mu^-$ and $B_s \to \mu^+\mu^-$, and big bang nucleosynthesis (BBN) are also shown. One can see that 
both the hidden monopole DM and the axion can be found by the future experiments such as PICO-500 and SHiP in
a parameter region around $m_a = {\cal O}(10)\,\text{MeV}$ and $f_a = {\cal O}(10^5)\,\text{GeV}$, and 
$m_a = {\cal O}(100)\,\text{MeV}$ and $f_a = {\cal O}(10^4)\,\text{GeV}$. The latter
 is opened  compared to the previous analysis~\cite{Dolan:2014ska}, and this is due to the hidden photon decay channel of the axion. Similar results would be obtained with other benchmark points.

\section{Discussion and Conclusions}\label{sec:5}
First, let us estimate the hidden photon contribution to dark radiation. In the minimal case without the axion portal coupling, the number of the extra neutrino species, $\Delta N_{\rm eff}$, was estimated in Ref.\,\cite{Khoze:2014woa}, where the entropy stored in the $W_{\rm H}^{\pm}$ is transferred to the hidden photons after the temperature falls below the mass of $W_{\rm H}^{\pm}$. Let us see how this is modified in our model setup. To be concrete, we fix the axion mass and decay constant to be the region of $m_a = {\cal O}(10)$\,MeV and $f_a = {\cal O}(10^5)$\,GeV, where both the hidden monopole DM and the axion are within the experimental reach. In this parameter region, the axion keeps the hidden photons in equilibrium with the SM sector at temperatures well below the mass of $W_{\rm H}^{\pm}$ or the PQ fermions, and the hidden photon is considered to decouple from the SM plasma at a temperature around the electroweak scale. Therefore, we do not need to take account of the entropy stored in the heavy hidden particles. For the reference parameter, the axion remains in equilibrium and abundant until the temperature becomes comparable to the axion mass. In particular, the axion mainly decays into the hidden photons when the axion becomes non-relativistic and its abundance gets Boltzmann suppressed. This production of the hidden photons dominates over the contribution at the decoupling of the hidden photons. Hence, we have $\Delta N_{\rm eff} \simeq 4/7 + 0.05 \sim 0.6$. In the other viable parameter region, $m_a = {\cal O}(100)$\,MeV, $f_a = {\cal O}(10^{4})$\,GeV, the contribution to dark radiation will be smaller because the axion disappears from the plasma at higher temperatures. The dark radiation with $\Delta N_{\rm eff} \gtrsim 0.5$ can relax the $H_0$ tension significantly~\cite{Riess:2019cxk}.\footnote{See Ref.~\cite{Takahashi:2019ypv} for the anthropic bound on  $\Delta N_{\rm eff}$.}

Next let us comment on a kinetic mixing between U(1)$_{\rm H}$ and the hypercharge U(1)$_Y$. As discussed in Refs.\,\cite{Brummer:2009cs,Bruemmer:2009ky}, we can write the following dimensional-5 operator
\begin{eqnarray}
\frac{1}{M} \left(\boldsymbol{\phi}  \cdot \boldsymbol{F}_{\rm H}^{\mu\nu}\right)  B_{\mu \nu} 
\,\supset\, \frac{\upsilon_{\rm H}}{M} F_3^{\mu \nu} B_{\mu \nu}
~, 
\end{eqnarray}
where $B_{\mu \nu}$ is the field strength tensor of U(1)$_Y$, and $M$ is a cut off scale. This effective operator gives rise to the kinetic mixing of $\upsilon_{\rm H}/M$, which is of order $10^{-13}$ for $\upsilon_{\rm H} = 10^5$\,GeV and $M = 10^{18}$\,GeV. Then, both monopoles and $W^\pm_{\rm H}$ acquire a fractional electric charge of order $\upsilon_{\rm H}/M$ through the kinetic mixing. The current bound on such mini-charged DM is satisfied for the mass of ${\cal O}(100)$\,TeV and the kinetic mixing of ${\cal O}(10^{-13})$~\cite{Kadota:2016tqq,Stebbins:2019xjr}.

In this paper, we have studied the possibility that the hidden monopoles which account for a sizable fraction of the observed DM density are coupled to the SM sector via the axion portal coupling. We have determined the axion configuration around the monopole so that it minimizes the energy of the system, and then estimated the scattering cross section of the monopole with a nucleon via the axion portal. Using the fact that the dependence of the cross section on the transferred momentum is the same as that for a CP-violating pseudo scalar coupling between quarks and a Dirac DM particle, we have translated the experimental bounds on the spin-dependent scattering cross section with a nucleon to limit the model parameters. The axion itself can also be searched for at the beam-dump experiment. Combining the current limits and future prospects for the search of DM and axions, we have found two viable regions around $m_a = {\cal O}(10)$\,MeV, $f_a = {\cal O}(10^{5})$\,GeV and $m_a = {\cal O}(100)$\,MeV, $f_a = {\cal O}(10^{4})$\,GeV where the monopole DM and the axion are respectively within the reach of the future experiments such as PICO-500 and SHiP.

\section*{Acknowledgments}
We thank Norimi Yokozaki for his collaboration in the early stage of the present work. This work is supported by JSPS KAKENHI Grant Numbers JP15H05889 (F.T.), JP15K21733 (F.T.),  JP17H02875 (F.T.), JP17H02878 (F.T.), 19J10946 (SY.H.), and by World Premier International Research Center Initiative (WPI Initiative), MEXT, Japan.

\end{document}